\documentclass[fleqn,usenatbib]{mnras}

\usepackage{newtxtext,newtxmath}
\usepackage{bm}
\usepackage{booktabs}
\usepackage{multirow}
\usepackage[utf8]{inputenc}

\usepackage[T1]{fontenc}

\DeclareRobustCommand{\VAN}[3]{#2}
\let\VANthebibliography\thebibliography
\def\thebibliography{\DeclareRobustCommand{\VAN}[3]{##3}\VANthebibliography}

\usepackage{graphicx}	
\usepackage{amsmath}	
\usepackage{amssymb}	
\usepackage{float}

\title[Probing magnetism in the cores of red giants]{Core magnetic field imprint in the non-radial oscillations of red giant stars}

\author[P. Gomes \& I. Lopes]{
Pedro Gomes,$^{1}$\thanks{E-mail: pedro.david.gomes@tecnico.ulisboa.pt}
Ilídio Lopes,$^{1}$\thanks{E-mail: ilidio.lopes@tecnico.ulisboa.pt}
\\
$^1$Centro de Astrof\'{\i}sica e Gravita\c c\~ao  - CENTRA, 
	Departamento de F\'{\i}sica \\ Instituto Superior T\'ecnico - IST,
	Universidade de Lisboa - UL \\ Av. Rovisco Pais 1, 1049-001 Lisboa
}

\date{Accepted XXX. Received YYY; in original form ZZZ}

\pubyear{2020}

\begin{document}
\label{firstpage}
\pagerange{\pageref{firstpage}--\pageref{lastpage}}
\maketitle

\begin{abstract}
Magnetic fields in red giant stars remain a poorly understood topic, particularly in what concerns their intensity in regions far below the surface. In this work, we propose that gravity-dominated mixed modes of high absolute radial order and low angular degree can be used to probe the magnetic field in their radiative cores. Using two poloidal, axisymmetric
configurations for the field in the core and the classical perturbative approach, we derive an analytical expression for the magnetic frequency splitting of these oscillation modes. Considering three distinct red giant models, with masses of 1.3\(M_\odot\), 1.6\(M_\odot\) and 2.0\(M_\odot\), we find that a field strength of $10^5$ G is necessary in the core of these stars to induce a frequency splitting of the order of a $\mu$Hz in dipole and quadrupole oscillation modes. Moreover, taking into account observational limits, we estimate that magnetic fields in the cores of red giants that do not present observable magnetic splittings cannot exceed $10^4$ G. Given the general absence of observable splittings in the oscillation spectra of these stars, and assuming that present mode suppression mechanisms are not biased towards certain azimuthal orders and
retain all peaks in each multiplet, our results lead us to conclude that internal fields with the considered configurations and strengths above $10^4$ G are not prevalent in red giants.
\end{abstract}




\section{Introduction}

Magnetic fields in stars remain one of the most enigmatic topics in stellar astrophysics. In the last 70 yr, powerful techniques such as the Zeeman effect \citep{1947ApJ...105..105B, 2009ARA&A..47..333D} and spectropolarimetry \citep{1997MNRAS.291..658D} have made possible the detection and measurement of magnetic activity in a plethora of stars. Of relevance are the detections in Ap and Bp stars, which host large-scale surface magnetic fields in the range $0.1 - 100$ kG \citep{ 2007A&A...475.1053A}, and A and OB-type stars, that possess small-scale, weak magnetic fields $(B \lesssim 100$G) \citep{2009A&A...500L..41L,2011A&A...534A.140C}, as this dichotomy raises important questions regarding the formation process of fossil fields \citep[e.g.][]{2017RSOS....460271B}. The aforementioned techniques, however, are limited to stellar surfaces, and therefore the issue of magnetism in the interior of stars remains mostly unexplored. In what concerns this uncertainty, asteroseismology has revealed itself capable of providing valuable insights and predictions.

In solar-like oscillators, turbulent convection in near-surface layers generates waves that propagate inwards throughout the star and interfere to construct an oscillation mode pattern \citep[e.g.,][]{2010aste.book.....A}. These modes can be acoustic, if the force responsible for restoring the star to equilibrium is pressure, or gravitational, if the restoring force is gravity (through buoyancy). Mixed oscillation modes also exist, possessing properties of both pressure and gravity modes. Although pressure modes have higher frequencies and live in the stellar envelope, often called the p-mode cavity, gravity modes have lower frequencies and can be used to obtain valuable insights about the stellar core, the g-mode cavity \citep[e.g.,][]{2013ARA&A..51..353C}. This sensitivity of the oscillation modes to the different regions inside stars is what allows the asteroseismological probing of stellar interiors at different depths, including the inference of magnetic field strengths and structure; for some time it has been known that weak magnetic fields lead to the splitting of the frequencies of
the oscillation modes \citep[e.g.,][]{ 1957AnAp...20..185L, 1982MNRAS.201..619B, 1983MNRAS.205.1171R}. This effect has been studied for the case of the Sun in what concerns acoustic and gravity modes \citep[e.g.][]{1984MmSAI..55..215G, 1990MNRAS.242...25G,2007MNRAS.377..453R}, Ap stars \citep{1985ApJ...296L..27D}, slowly pulsating B-stars \citep{2005A&A...444L..29H}, $\beta$ Cephei stars \citep{2000ApJ...531L.143S} and white dwarfs \citep{1989ApJ...336..403J}.

\begin{figure*}
    \centering
    \begin{minipage}[b]{\columnwidth}
        \centering
        \includegraphics[scale=0.5]{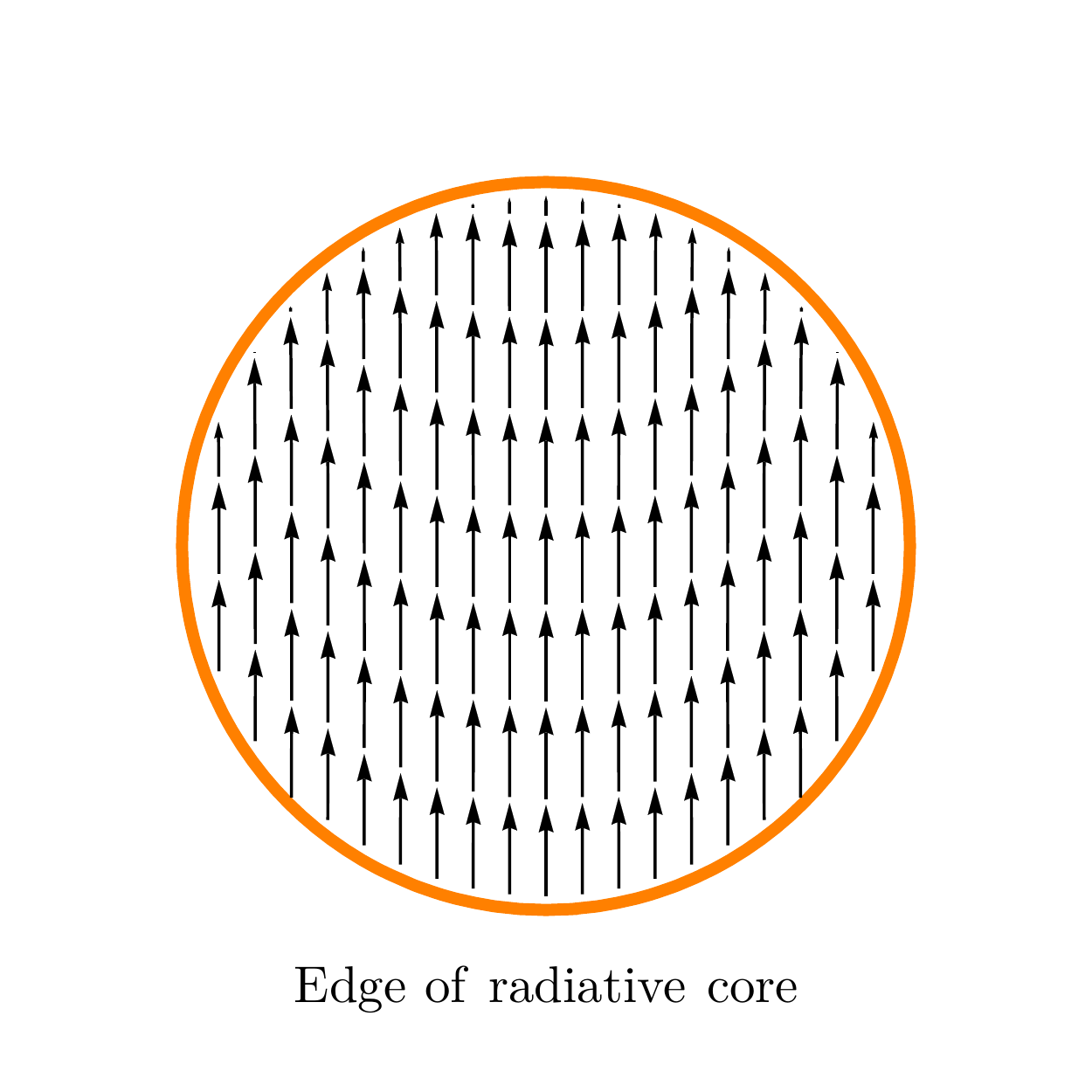}
    \end{minipage} \hfill
    \begin{minipage}[b]{\columnwidth}
        \centering
        \includegraphics[scale=0.5]{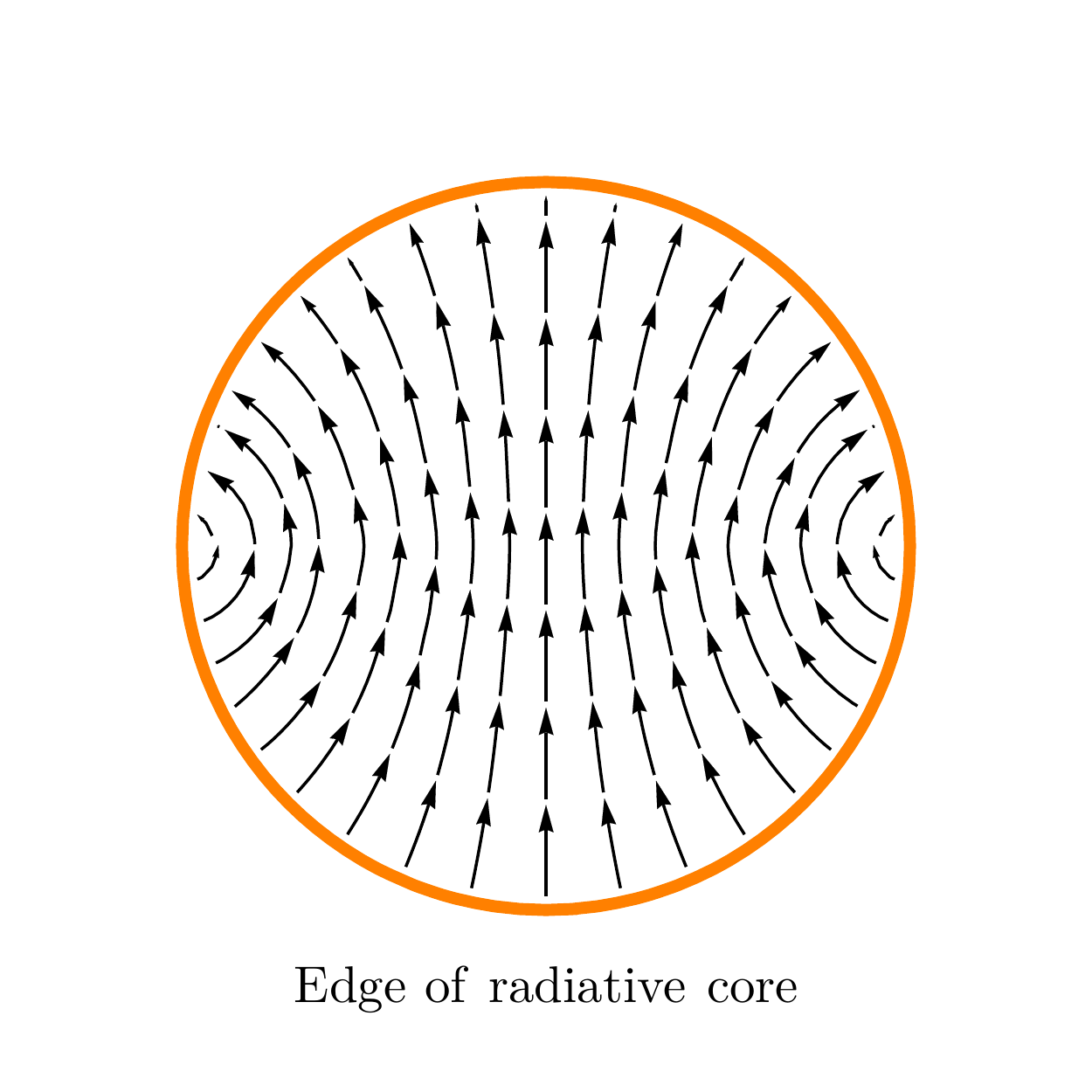}
    \end{minipage} \vspace{0.3cm}
    \caption[]{Illustration of the magnetic field lines of the poloidal field configuration proposed by \protect\cite{1981AN....302...65R}, expression \ref{field1} (left figure) and by \protect\cite{2004physics...9093K}, expression \ref{field2} (right figure) throughout the stellar radiative core, delimited in orange.}
    \label{magnetic1}
\end{figure*}

In the recent years, the sheer wealth and amount of data collected by the \textit{CoRoT} \citep{2007AIPC..895..201B} and Kepler missions \citep[e.g.][]{2010PASP..122..131G,2011Sci...332..213C,2017ApJS..233...23S} has led to an enormous up-growth of knowledge regarding the structure and constitution of stars and allowed the testing of theories of stellar evolution and seismology. This has certainly been the case in what concerns red giant stars. Following the cessation of nuclear reactions in the core at the end of main-sequence (MS) phase, the stellar envelope expands, whereas the core contracts and the star ascends the red giant branch (RGB). This duality leads the acoustic modes in the convective envelope to couple with the gravity modes in the radiative core, forming a very rich and complex mixed-mode pattern that can be used to probe the star in all its extension \citep[e.g.,][]{2013ARA&A..51..353C}. Mixed modes in red giants have proven to be extremely useful in measuring core rotation rates \citep[e.g.][]{2012Natur.481...55B}, distinguishing between stars undergoing hydrogen or helium burning \citep{2011Natur.471..608B} and understanding the importance of rotation and convective overshooting in the core via period spacings \citep{2016MNRAS.457L..59L}. Recently, it has also been proposed that strong magnetic fields can be detected in the cores of red giants showing depressed oscillation modes \citep{2015Sci...350..423F,2016ApJ...824...14C} although such conclusions were called into question by \cite{2017A&A...598A..62M}. The processes of field generation and conservation in the interior of these stars are still largely unexplored, with only a few works discussing dynamo mechanisms in the envelope to explain isotopic abundances in their atmospheres \citep{2007ApJ...671..802B,2008ApJ...684L..29N} and some theories that resort to fossilised fields to explain magnetism in their radiative cores \citep[e.g][]{2016ApJ...824...14C}. Confirmed detection of magnetic activity in red giants is, such as the majority of stars, limited to their surfaces, with several dozens exhibiting magnetic fields detected by spectropolarimetry, ranging from the sub-Gauss to the tenths of Gauss \citep[]{2014IAUS..302..373K, 2015A&A...574A..90A}.

Studies of the character of mixed modes in red giants reveal that a high number of these stars have mixed modes dominated by gravity \citep[e.g][]{2009MNRAS.400L..80S}. Hence, they behave almost as pure gravity modes and can be used to probe the physical conditions in the stellar core. With this in mind, we derive an expression for the magnetic splitting of gravity-dominated (g-dominated) mixed modes in red giants by a weak core magnetic field. Using this expression, we estimate the core field strengths that would lead to a measurable frequency splitting in the oscillation spectra. This work is organised as follows $-$ in Section \ref{sec:splitting} we start with the general discussion of magnetic frequency splittings, and elaborate on the chosen field configurations used in our treatment of the problem, based on the numerical simulations performed by  \cite{2007A&A...469..275B}. We then derive the aforementioned expression, which is similar to the one found by \cite{2005A&A...444L..29H} and \cite{2007MNRAS.377..453R}. We also briefly address rotational splittings, and end this section with a discussion of the limitations of our approach. In Section \ref{sec:model}, we explain the chosen stellar models and present the results for the core magnetic field strength that would produce observable frequency splittings, sufficiently distinguishable from the ones induced by rotation, for dipole and quadrupole mixed oscillation modes. Moreover, we proceed to do the same for a splitting below the frequency resolution, thus establishing constraints on the core magnetic field strength of red giants that do not show magnetic splittings in their oscillation spectra. Lastly, in Section \ref{sec:discussion}, we discuss these results.

\section{Splitting of the oscillation modes}
\label{sec:splitting}
\subsection{Splitting by a magnetic field}

The presence of a magnetic field can modify the oscillation mode pattern by introducing a magnetic frequency splitting. If the magnetic field is weak enough that the splitting is small in comparison with the spacing between unperturbed modes and that it does not significantly affect the fluid motions, a perturbative treatment is justifiable, and, in these conditions \citep[e.g.,][]{1989nos..book.....U} the splitting $\delta \omega$ is

\begin{equation}
\begin{split}
    \frac{\delta \omega}{\omega} = - \frac{1}{8 \pi \omega^2 I} \int \big[ & \big( \bm{\nabla} \bm{\times} \boldsymbol{B}'\big) \bm{\times} \boldsymbol{B} + \big(\bm{\nabla} \bm{\times} \boldsymbol{B} \big) \bm{\times} \boldsymbol{B}' \\ & + \frac{\rho'}{\rho} \boldsymbol{B} \times \big( \nabla \times \boldsymbol{B} \big) \big] \cdot \boldsymbol{\xi}^* \, \text{d}V \, . \label{splitting}
    \end{split}
\end{equation}

In this expression, $\omega$ is the angular frequency of the mode, $\boldsymbol{B}$ is the magnetic field vector, $\rho$ is the density, and the same primed symbols denote their eulerian perturbations. In turn, $\bm{\xi}$ is the displacement vector in the nonmagnetic unrotating star, given by, in spherical coordinates $(r, \theta, \phi)$,

\begin{equation}
    \bm{\xi}(r, \theta, \phi, t) = \left( \xi_r(r) Y^\ell_m , \xi_h(r) \frac{\partial Y^\ell_m}{\partial \theta} , \xi_h(r) \frac{i m}{\sin \theta} Y^\ell_m \right) \text{e}^{i \omega t}
    \label{displacement}
\end{equation}

if torsional motions are not considered, where $\xi_r(r)$ and $\xi_h(r)$ are the radial and horizontal displacement functions, respectively. As usual, $Y^\ell_m$ are the spherical harmonics functions of angular degree $\ell$ and azimuthal order $m$, and $t$ is the time. In turn, $I$ is the mode inertia:
\begin{equation}
    I = \int \left( \xi_r^2 + \ell(\ell+1) \xi_h^2 \right) \rho \, r^2 \, \text{d}r.
    \label{mode_inertia}
\end{equation}

Assuming a stationary magnetic field in the absence of magnetic diffusion, the perturbation to the magnetic field, $\boldsymbol{B}'$, can be obtained from the induction equation:

\begin{equation}
    \boldsymbol{B}' (r, \theta, \phi, t) = \bm{\nabla} \bm{\times} \left( \bm{\xi} \bm{\times} \boldsymbol{B} \right) \label{perturbed} \, .
\end{equation}

Using this expression, the first term inside the integral in expression \ref{splitting} can be written as \citep[e.g.,][]{1989nos..book.....U}

\begin{equation}
    - \left[ \left( \bm{\nabla} \bm{\times} \boldsymbol{B}' \right) \bm{\times} \boldsymbol{B} \right] \cdot \bm{\xi}^* = |\boldsymbol{B}'|^2. \label{absfield}
\end{equation}

In all subsequent algebra, we omit the factor $\text{e}^{i \omega t}$. Although the perturbed magnetic field $\boldsymbol{B}'$ is time-dependent, we are interested in computing $|\boldsymbol{B}'|^2$, and since $|\text{e}^{i \omega t}| = 1$ we opt to omit this factor.

\subsection{Magnetic field configurations}
\label{subsec:config}

In this work, we consider two poloidal configurations aligned with the axis of rotation for the magnetic field in the radiative core. We make no assumptions regarding the formation or conservation process of this field, we simply assume it exists and that the chosen configurations are representative of its poloidal component. The first was proposed by  \cite{1981AN....302...65R} and, in the spherical coordinate system, takes the form

\begin{equation}
    \boldsymbol{B}_1 (r, \theta, \phi) = \frac{B_{0}}{2} \Big( \cos \theta \, (5-3x^2), \sin \theta \, (6x^2-5), 0 \Big) \label{field1}
\end{equation}

where $B_{0}$ is a constant term, $x = r/R$ is the normalised radial coordinate and $R$ is the stellar radius. This field is illustrated on the left side of Fig. \ref{magnetic1} and, because of its resemblance to a uniform magnetic field, was considered for its simplicity. The second field configuration was proposed by \cite{2004physics...9093K},
\begin{equation}
     \boldsymbol{B}_2 (r, \theta, \phi) = B_{0} \Bigg( \frac{\cos \theta}{(1+y^2)^2}, \frac{\sin \theta (y^2-1)}{(1+y^2)^3}, 0 \Bigg) \label{field2}
\end{equation}

where $y = x/c$, $c$ being a constant. The field lines for this configuration are shown on the right of Fig. \ref{magnetic1}.

The choice of these two fields was motivated by the numerical analysis carried out by \cite{2007A&A...469..275B} in what concerns hydrodynamical stability in the presence of rotation. The author concluded that these field configurations were always unstable, regardless of the angular velocity of rotation and the angle between the rotation axis and the axis of the magnetic field. The instability of purely poloidal or toroidal fields had already been demonstrated analytically \citep[e.g.][]{1973MNRAS.161..365T,1973MNRAS.163...77M,1973MNRAS.162..339W} and reinforced by numerical simulations. However, these simulations also reveal that a dynamical stable state can be achieved if both components exist and have similar strengths \citep[e.g.,][]{2004Natur.431..819B,2006A&A...450.1077B}. According to Braithwaite, the second field falls in this category and can be stable (in the absence of rotation) if an adequate toroidal component is considered. In other words, a mixed configuration involving this poloidal component is plausible. Although relatively simple and posterior mixed configurations have been proposed \citep[e.g.,][]{2010A&A...517A..58D,2010ApJ...724L..34D,2017MNRAS.467.3212L}, including very recent ones \citep[e.g.,][]{2019A&A...627A..64P,2020A&A...636A.100P,2020MNRAS.493.5726L} they generally demonstrate a functional similarity with those used in this work, for instance, in what concerns the angular dependence of the radial component with $\cos \theta$ and of the latitudinal component with $\sin \theta$. Besides, even if we were to consider a mixed configuration for the sake of stability, our results would not change greatly, as the toroidal component makes little contribution to the splitting in comparison with the poloidal counterpart (which is explained in the following section). The choice of these two fields also has the advantage that it leads to a relatively simple analytical expression for the magnetic frequency splitting.

We make no assumptions regarding the magnetic field in the stellar envelope, for two reasons: first, because red giants possess convective envelopes, and therefore the study of the field in this region must be based on a dynamo theory, something that is not within the scope of this work. Secondly, and most importantly, our calculations are performed for red giants in more advanced stages of evolution in the RGB, for which the g-dominated mixed modes possess most of their inertia in the core (this can be seen for one of the stellar models in Section \ref{sec:model}) and are not strongly affected by the physical conditions in the envelope. The whole reasoning carried out in this paper would not be valid, for instance, for red giants at the base of the RGB, where a significant fraction of the mode inertia still resides in the envelope. Thus, without loss of generality, we take the field in the envelope to be zero, and consider only the field in the core for our calculations \ref{field1} and \ref{field2}).

\subsection{Gravity-dominated mixed modes}
\label{sec:asymptotic_expansion}

In this work, we use the general expression for $\boldsymbol{B}'$ presented by \cite{2007MNRAS.377..453R}, which can be found in the Appendix, expression \ref{pertfield}. For the two considered fields, each of the components of the perturbed magnetic field can also be found in the Appendix,  expressions \ref{bpr} - \ref{bpphi}, as well as a more detailed explanation on the approximations that follow.

Expression \ref{splitting} can be simplified by taking into account that, in a large number of red giant stars, the mixed oscillation modes are dominated by gravity \citep[e.g][]{2009MNRAS.400L..80S}. These modes have, in the standard Eckart-Scuflaire-Osaki-Takata scheme \citep{1960hydro.book.....F,1974A&A....36..107S,1975PASJ...27..237O,2006PASJ...58..893T}, high absolute radial orders $|n| \equiv |n_p - n_g|$, where $n_p$ and $n_g$ are the number of nodes of the radial displacement function $\xi_r$ in the regions where the mode behaves like a p-mode or a g-mode, respectively. For modes of high $|n| $ and angular frequency $\omega \ll N, S_\ell$, where $N$ is the Brunt–Väisälä frequency and $S_\ell$ is the Lamb frequency for a mode with angular degree $\ell$, the JWKB approximation is valid and the horizontal and radial displacement functions can be written, respectively \citep[e.g.][]{1989nos..book.....U,2010aste.book.....A} 
\begin{equation}
\begin{split}
  \xi_h(r) \simeq & - A \, \rho^{-1/2} \, r^{-3/2} \, \left[ \ell (\ell + 1 ) \right]^{-1} \left( \frac{N^2}{\omega^2} - 1 \right)^{1/4} \\
  & \sin \left[ \int_{r_1}^r \frac{\ell(\ell+1)}{r} \left( \frac{N^2}{\omega^2} - 1 \right)^{1/2} \, dr' - \frac{\pi}{4} \right]
\end{split}
\end{equation}
\begin{equation}
\begin{split}
  \xi_r(r) \simeq & \, A \, \rho^{-1/2} \, r^{-3/2} \, \left( \frac{N^2}{\omega^2} - 1 \right)^{-1/4} \\
  & \cos \left[ \int_{r_1}^r \frac{\ell(\ell+1)}{r} \left( \frac{N^2}{\omega^2} - 1 \right)^{1/2} \, dr' - \frac{\pi}{4} \right]
\end{split}
\end{equation}

where $A$ is a constant and $r_1$ is the lower bound of the g-mode cavity. The ratio of the amplitude of the horizontal displacement to the radial displacement is

\begin{equation}
    \left| \frac{\ell(\ell+1) \xi_h}{\xi_r} \right| \sim \left( \frac{N^2}{\omega^2} - 1 \right)^{1/2}
\end{equation}

which is much greater than unity throughout most of the radiative core, where $N \gg \omega$. Hence, since $|\xi_h| \gg |\xi_r|$, the oscillations occur mostly in the horizontal direction, and the mode inertia, expression \ref{mode_inertia}, can be simplified:
\begin{equation}
    I \simeq \ell(\ell+1) \int \xi_h^2 \, \rho \, r^2 \, dr
    \label{i_simplified}
\end{equation}

In turn, since the modes have very high absolute radial orders $|n|$, the terms involving the derivatives of the horizontal displacement, $\partial \xi_h / \partial r$, are very large in the core in comparison with terms involving $\xi_h$ and $\xi_r$. This can be seen by differentiating $\xi_h$ with respect to the radial coordinate, which leads to the emergence of the factor $(N^2/\omega^2 - 1)^{1/2} \gg 1$, that makes such terms dominant. Because of this property of the modes, expression \ref{absfield}, that corresponds to the first term in the integral of expression \ref{splitting}, can be simplified to

\begin{equation}
\begin{split}
  |\mathbf{B}'|^2 \simeq & \left( \frac{B_0}{R} \right)^2 \left| \frac{1}{x} \frac{\partial }{\partial x} (x \, f_i(x) \, \xi_h) \right|^2 \Bigg( \left| \cos \theta \frac{\partial Y^\ell_m}{\partial \theta} \right|^2 \\ & + m^2 \left| \cot \theta \, Y^\ell_m \right|^2  \Bigg)
\end{split}
\label{eqn1}
\end{equation}

where, once more, $B_0$ is a constant, $R$ is the stellar radius, $x$ is the normalised radial coordinate, $f_1(x) = (5-3x^2)/2$ for the field given by expression \ref{field1} and $f_2(x) = 1/(1+(x/c)^2)^2$ for the field given by \ref{field2}. In the integral of expression \ref{splitting}, this first term is dominant in comparison with the remaining, as it is quadratic in the derivatives of $\xi_h$ whereas the second scales with $\partial \xi_h / \partial x$ and the third with $\xi_h$. Hence, expression \ref{splitting} reads

\begin{equation}
    \frac{\delta \omega}{\omega} = \frac{1}{8 \pi \omega^2} \left( \frac{B_0}{R} \right)^2 \, \mathcal{I} \, C_{\ell,m} \label{splittingsimple}
\end{equation}

where

\begin{equation}
    \mathcal{I} = \frac{\int \left| \frac{1}{x} \frac{\partial }{\partial x} (x \, f_i(x) \, \xi_h) \right|^2 \, x^2 dx}{ \ell(\ell+1) \int \xi_h^2 \, \rho \, x^2 \, \text{d}x} \label{i}
\end{equation}

with $i = 1,2$ concerning the fields \ref{field1} and \ref{field2}, respectively, and the coefficients $C_{\ell,m}$ are
\begin{equation}
    C_{\ell,m} = \int \left( \left| \cos \theta \frac{\partial Y^\ell_m}{\partial \theta} \right|^2 + m^2 \left| \cot \theta \, Y^\ell_m \right|^2 \right) \, \sin \theta \, d\theta. \label{coefficients}
\end{equation}

Given the insensitivity of these coefficients to the sign of the azimuthal order $m$, the effect of the field is to lift the degeneracy and produce $\ell + 1$ eigenmodes \citep{1957AnAp...20..185L}. This result is, of course, only relevant for non-radial modes; since the azimuthal order varies from $-\ell \leq m \leq \ell$, and for radial modes one has $\ell = 0$, then $\delta \omega = 0$ and there is no degeneracy. Expression \ref{splittingsimple} and the following are similar to the ones found by \cite{2005A&A...444L..29H} and \cite{2007MNRAS.377..453R} for the splitting of the frequencies of g-modes in a B-star model and in the Sun, respectively. 

As we mentioned in Section \ref{subsec:config}, although magnetic fields must have both poloidal and toroidal components to be stable, the toroidal component would cause a much smaller splitting than the poloidal one. A toroidal magnetic field aligned with the rotation axis has only one nonzero component, the azimuthal one, and using expression \ref{pertfield} in the Appendix for $\boldsymbol{B}'$ it is straightforward to see that for such fields the dominant term in the integral of expression \ref{splitting} is $|\xi_h|^2$. This term is much smaller than the term involving $|\partial \xi_h/\partial x|^2$, that emerges from the poloidal component, and negligible by comparison. Therefore, we do not expect the toroidal counterpart of the field to make a significant contribution to the splitting and we can safely discard it.

\subsection{Splitting by rotation}
\label{sec:rotation}

The presence of rotation also leads to the splitting of the frequencies of the oscillation modes. Assuming that the angular frequency of rotation $\Omega$ is only a function of the radial coordinate, $\Omega = \Omega(r)$, the rotational splitting is \citep[e.g.,][]{2010aste.book.....A}
\begin{equation}
    \delta \omega = m \int_0^R K_{n\ell} (r) \, \Omega(r) \, dr
\end{equation}

where $K_{n\ell}$ is the rotational kernel
\begin{equation}
    K_{n\ell} = \frac{\left[ \xi_r^2 + \ell(\ell+1)\xi_h^2 -2 \xi_r \xi_h -\xi_h^2 \right] r^2 \rho}{\int_0^R \left[ \xi_r^2 + \ell(\ell+1) \xi_h^2 \right] r^2 \rho \, dr} . \label{kernel}
\end{equation}

In the case of uniform rotation, the rotational splitting simplifies to $\delta \omega = m \beta_{n \ell} \, \Omega$, where $\beta_{n \ell}$ is the integral of the rotational kernel over the radius of the star. Separating the contributions to the splitting from the core and the envelope and assuming that the former rotates much faster than the latter (thus neglecting the contribution of the envelope to the splitting of the frequencies), the rotational splitting for g-dominated mixed modes is \citep{2012A&A...548A..10M,2013A&A...549A..75G}

\begin{equation}
    \delta \omega \simeq m \left< \Omega_\text{c} \right> \left( 1 - \frac{1}{\ell(\ell+1)} \right) \zeta \, ,
\end{equation}

where $\left< \Omega_\text{c} \right>$ is the kernel-averaged angular velocity in the core

\begin{equation}
    \left< \Omega_\text{c} \right> = \frac{\int_0^{r_c} \Omega(r) K_{n\ell} (r) \, dr }{\int_0^{r_c} K_{n\ell} (r) \, dr}
\end{equation}

$r_c$ is the radius of the core, and $\zeta$ is the ratio of the mode inertia in the core to the total mode inertia, $\zeta = I_\text{c}/I \simeq 1$. Hence, if the core is rigidly rotating, the splitting of g-dominated modes is roughly of the same order of magnitude as the core angular velocity. Rotational splittings have been measured in a plethora of red giants, and they typically range between the hundredths and tenths of $\mu$Hz \cite[e.g.][]{2012A&A...548A..10M}.

\subsection{Limitations}
\label{subsec:limit}

In the derivation of expression \ref{splittingsimple} for the splitting of the eigenfrequencies of red giant stars by a magnetic field in the core, a set of assumptions was made that needs explanation with regard to their limitations:

\begin{enumerate}[i]

\item Firstly, as mentioned in the first paragraph of Section \ref{sec:splitting}, expression \ref{splitting} holds when the effect of the magnetic field on the frequencies is to cause a splitting that is small when compared to the spacing between unperturbed modes. If this is not the case, i.e., if the perturbation is large enough that the frequency of the perturbed mode approaches an adjacent unperturbed mode, then the latter needs to be included in the calculation of the splittings, as it can significantly contribute to the perturbed mode. In this event, quasi-degenerate perturbation theory should be used. This has been studied by  \cite{2017ApJ...846..162K,2018ApJ...854...74K}, who, using an ansatz derived by \cite{1992RSPTA.339..431L} for calculating the coupling strength between oscillation modes in the context of quasi-degenerate perturbation theory, have
obtained an analytical expression for the coupling strength in the form of a general matrix element. According to the authors, this matrix element can, in turn, be used to construct a supermatrix whose eigenvalues consist on the perturbed frequencies, and the eigenvectors can be used to obtain the perturbed eigenfunctions. In resume, expression \ref{splitting} is obtained by using a standard, linear perturbation analysis, which may break down for a large number of stars. In those cases, it is necessary to consider a generalized perturbation approach, for which the coupling of the modes can not be disregarded.

\begin{figure}
\medskip
\centering
\includegraphics[width=8 cm,keepaspectratio]{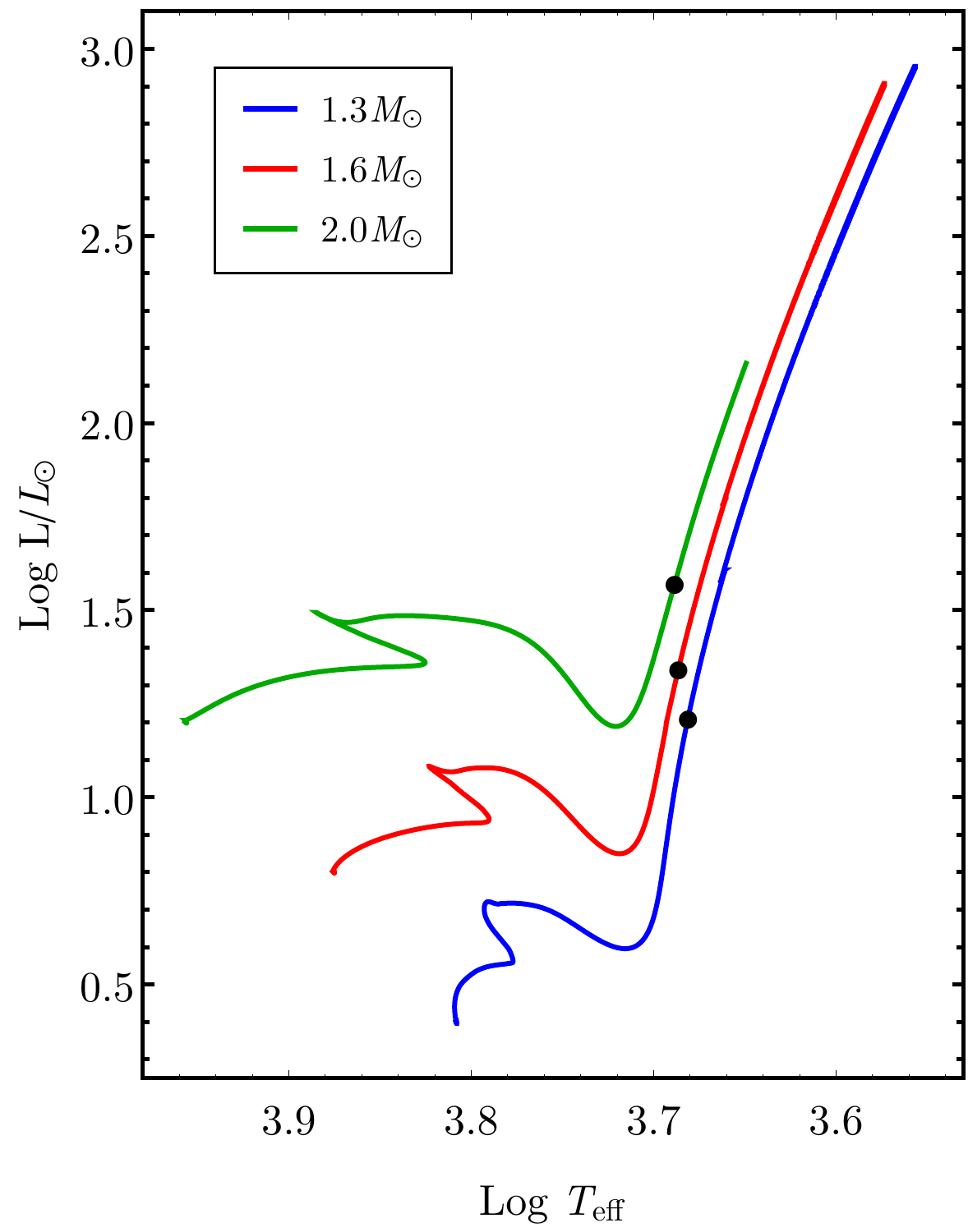}
\caption{HR diagram for 1.3\(M_\odot\), 1.6\(M_\odot\) and 2.0\(M_\odot\) stars (blue, red and green curves, respectively) from the MS to the tip of the RGB. The black dots indicate the chosen moments during red giant evolution to study the splitting of the frequencies of the oscillation modes by a magnetic field in the core. For the 1.3\(M_\odot\) this corresponds to an age of 4.522 Gyr, and to the 1.6\(M_\odot\) and 2.0\(M_\odot\) to an age of 2.196 and 1.109 Gyr, respectively.} 
\label{hrdiagram}
\end{figure}

\item Secondly, as also mentioned in the first paragraph of Section \ref{sec:splitting}, we assume that the magnetic field does not severely impact the motion of the fluid. As was suggested by \cite{2015Sci...350..423F} and further developed by \cite{2017MNRAS.466.2181L} and \cite{2018MNRAS.477.5338L}, magnetic fields may significantly influence gravity wave propagation or even disrupt it. These authors concluded that magnetic fields with strengths above a critical value, not strong enough to alter the stellar structure, can lead to the transfer of energy of gravity waves into magnetic waves, which eventually dissipate once they propagate to regions where the magnetic field is weaker. Our results for the field strength that can cause an observable frequency splitting, presented in Section \ref{sec:model}, are very close or may even be on the verge of exceeding this critical field strength for which there is this regime change. Should this be the case, the assumption that the perturbed eigenfunction is small when compared to the unperturbed eigenfunction is no longer true, and perturbation theory breaks down. Thus, there is the need to assume that the magnetic field does not impact gravity wave propagation significantly and that, in this context, expression \ref{splitting} holds.

\item Thirdly, there is the assumption that the effects of the magnetic field do not have to be accounted for in the stellar structure. Since expression \ref{splitting} is obtained in the context of a perturbative analysis, one of the starting points is that the magnetic field does not significantly alter the structure of the star and that the equilibrium state depends on the balance between pressure gradients and gravity. In the case of red giant stars, this is a realistic assumption, as the contraction of the radiative core throughout the evolution leads to extremely high central pressures, and the pressure gradients are orders of magnitude above the Lorentz force; this was verified by considering typical field strengths of  $10^5$ G, motivated by our results discussed next. However, if this were not the case, it would be necessary to take into account the effects of the magnetic field on the equilibrium model.

\end{enumerate}

Due to these limitations, our results are only expected to hold when the magnetic field is relatively weak (below the critical field strength mentioned in the second point). Nevertheless, as is reinforced multiple times in this paper, the purpose of the presented results and the subsequent discussion is to provide possible estimates for the order of magnitude of magnetic field strengths in the cores of red giant stars.

\begin{figure}
\medskip
\centering
\includegraphics[width=8 cm,keepaspectratio]{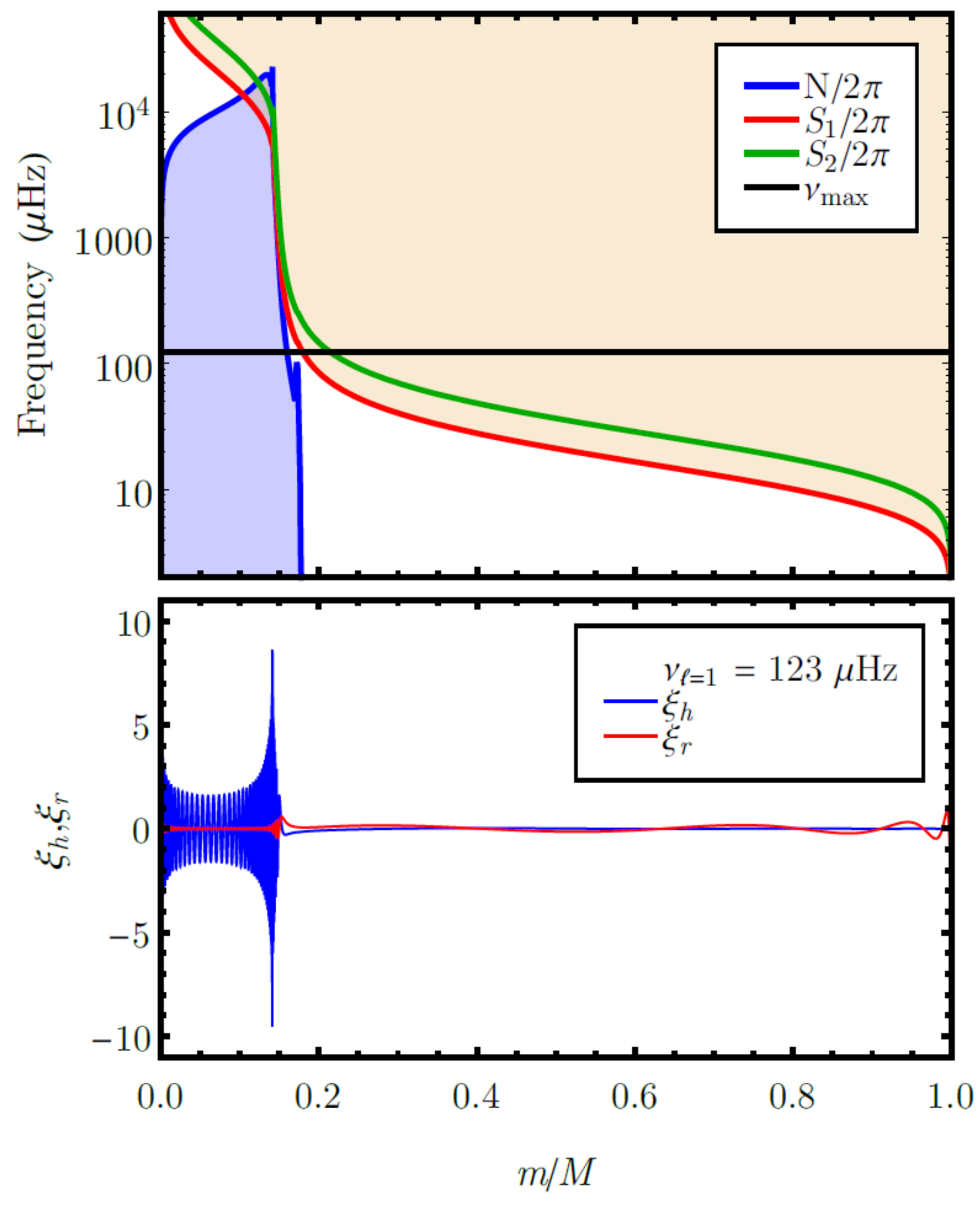}
\caption{Top: propagation diagram for a red giant model with an initial mass of 1.6\(M_\odot\), at an age of 2.196 Gyr and estimated $\nu_\text{max} = 123 \, \mu$Hz as a function of the normalised stellar mass. The blue curve is the Brunt–Väisälä frequency and the red and green curves are the Lamb frequency for dipole ($\ell = 1$) and quadrupole ($\ell = 2$) modes, respectively. The horizontal black line is the frequency of maximum power $\nu_\text{max}$, estimated from scaling relations. The blue region is the g-mode cavity and the red region is the p-mode cavity for dipole modes. Bottom: scaled horizontal and radial displacement eigenfunctions, blue and red curves, respectively, for a dipole mode with $\nu = 123 \, \mu$Hz and absolute radial order $|n| = 98$.} 
\label{propagation}
\end{figure}

\section{The stellar models}
\label{sec:model}

The stellar models considered in this work were constructed using MESA  (Modules for Experiments in Stellar Astrophysics), release 10398 \citep{2011ApJS..192....3P,2013ApJS..208....4P,2015ApJS..220...15P,2018ApJS..234...34P,2019ApJS..243...10P}. We used the prescription of \cite{2015Sci...350..423F} and evolved three red giant models, with masses 1.3\(M_\odot\), 1.6\(M_\odot\) and 2.0\(M_\odot\), from the MS to the RGB. In Fig. \ref{hrdiagram} we show the respective HR diagram, where the black dots indicate the chosen moments in the models’ evolution to study the splitting of the frequencies of the oscillation modes by a magnetic field in the core. These values for the masses were chosen to cover the range of masses of the red giant population observed by the Kepler space mission. To obtain the frequencies of the oscillation modes, we started by estimating the frequency of maximum power $\nu_\text{max}$ using well-known scaling relations \citep[e.g.,][]{2011A&A...530A.142B}. Afterwards, we proceeded to compute the eigenfrequencies closest to these values, using GYRE 5.1. \citep{2013MNRAS.435.3406T} in the adiabatic regime with the default boundary conditions. We chose three distinct frequencies for our analysis, the middle-valued one being the closest to the estimated $\nu_\text{max}$.

\begin{table}
\centering
        \begin{tabular}{c c c c c c}
            \toprule
             &  &  & \multicolumn{2}{c}{$C_{1,|1|} - C_{1,0}$} \\ \cline{4-5}
             & $|n|$ & $\nu \, (\mu$Hz) & $B_\text{core,1}$ (MG) & $B_\text{core,2}$ (MG) \\ \hline
            \multirow{3}{*}{$1.3$\(M_\odot\)} 
                                 & 100 & 118.41 & 0.258 & 0.259 \\
                                 & 89 & 130.16 & 0.298 & 0.299  \\ 
                                 & 80 & 141.40 & 0.337 & 0.339 \\ 
                                 \hline
            \multirow{3}{*}{$1.6$\(M_\odot\)} 
                                 & 108 & 113.66 & 0.241 & 0.243  \\ 
                                 & 98 & 123.00 & 0.269 & 0.270  \\ 
                                 & 88 & 133.89 & 0.331 & 0.332 \\
                                 \hline
            \multirow{3}{*}{$2.0$\(M_\odot\)} 
                                 & 90 & 84.77 & 0.496 & 0.498  \\ 
                                 & 80 & 93.07 & 0.570 & 0.572  \\ 
                                 & 70 & 102.71 & 0.669 & 0.670 \\
                                 \hline
        \end{tabular}
    \caption{Value of the magnetic field strength in the core of 1.3\(M_\odot\), 1.6\(M_\odot\) and 2 \(M_\odot\) red giant models necessary to produce a frequency splitting of $1 \, \mu$Hz in the frequencies $\nu$ of mixed dipole ($\ell = 1$) modes with absolute radial order $|n|$, for the chosen field configurations. $B_\text{core,1}$ concerns the first field, given by expression \ref{field1}, whereas $B_\text{core,2}$ concerns the field given by expression \ref{field2}.}
    \label{tabledipole}
\end{table}

\begin{table*}
\centering
        \begin{tabular}{c c c c c c c c c c c}
            \toprule
             &  &  & \multicolumn{2}{c}{$C_{2,0} - C_{2,|1|}$} & & \multicolumn{2}{c}{$C_{2,0} - C_{2,|2|}$} & & \multicolumn{2}{c}{$C_{2,|1|} - C_{2,|2|}$} \\ \cline{4-5} \cline{7-8} \cline{10-11} 
             & $|n|$ & $\nu \, (\mu$Hz) & $B_\text{core,1}$ (MG) &  $B_\text{core,2}$ (MG) & & $B_\text{core,1}$ (MG) &  $B_\text{core,2}$ (MG) & & $B_\text{core,1}$ (MG) &  $B_\text{core,2}$ (MG) \\ \hline 
            \multirow{3}{*}{$1.3$\(M_\odot\)} 
& 180 & 118.57 & 0.305 & 0.307 & & 0.153 & 0.154 & & 0.176 & 0.177 \\
& 162 & 130.12 & 0.352 & 0.355 & & 0.176 & 0.178 & & 0.203 & 0.205 \\
& 147 & 141.47 & 0.399 & 0.402 & & 0.200 & 0.201 & & 0.231 & 0.232 \\ \hline
            \multirow{3}{*}{$1.6$\(M_\odot\)} 
& 196 & 112.69 & 0.274 & 0.276  & & 0.137 & 0.138 & & 0.158 & 0.159 \\  
& 177 & 123.39 & 0.313 & 0.314 & & 0.156 & 0.157 & & 0.180 & 0.181 \\
& 162 & 133.24 & 0.351 & 0.353 & & 0.175 & 0.176 & & 0.203 & 0.204 \\
\hline
            \multirow{3}{*}{$2.0$\(M_\odot\)} 
& 163 & 84.75 & 0.583 & 0.585 & & 0.292 & 0.293 & & 0.337 & 0.338 \\ 
& 147 & 92.72 & 0.673 & 0.676 & & 0.337 & 0.338 & & 0.389 & 0.390 \\ 
& 132 & 101.65 & 0.766 & 0.768 & & 0.383 & 0.384 & & 0.442 & 0.443 \\ \hline
        \end{tabular}
    \caption{Value of the magnetic field strength in the core of 1.3\(M_\odot\), 1.6\(M_\odot\) and 2 \(M_\odot\) red giant models necessary to produce a frequency splitting of $1 \, \mu$Hz in the frequencies $\nu$ of mixed quadrupole ($\ell = 2$) modes with absolute radial order $|n|$, for the chosen field configurations. $B_\text{core,1}$ concerns the first field, given by expression \ref{field1}, whereas $B_\text{core,2}$ concerns the field given by expression \ref{field2}.}
    \label{tablequadropole}
\end{table*}

In the top plot of Fig. \ref{propagation} we show the propagation diagram in normalized mass coordinates for the 1.6\(M_\odot\) model, at an age of 2.196 Gyr and with an estimated $\nu_\text{max} = 123 \, \mu$Hz (horizontal black line). The blue curve is the Brunt–Väisälä frequency and the red and green curves are the Lamb frequency for dipole $(\ell = 1)$ and quadrupole modes $(\ell = 2)$, respectively. The blue-shaded region corresponds to the g-mode cavity, where the oscillations are dominated by gravity. This is evident in the bottom plot, where the horizontal and radial displacements are shown (the blue and red curves, respectively) for the dipole oscillation mode with frequency closest to the estimated $\nu_\text{max}$ and absolute radial order $|n| = 98$. As can be seen, throughout the radiative core $|\xi_h| \gg |\xi_r|$, and, given the high absolute radial order, $|\partial \xi_h / \partial x| \gg |\xi_h|, |\xi_r|$, which were the base assumptions for the derivation of expression \ref{splittingsimple}. This type of reasoning, of course, holds for the other two models as well.

As mentioned in Section \ref{sec:rotation}, rotational splittings in red giants are typically in the range $10^{-2} - 10^{-1}$ $\mu$Hz \citep{2012A&A...548A..10M}. With these values in mind, we determined the necessary magnetic field strength in the core to induce a 1 $\mu$Hz splitting between modes with the same angular degree $\ell$ and different azimuthal order $m$, sufficiently distinguishable from the splitting caused by rotation. Although the mixed-mode pattern in red giants is intricate and the task of disentangling rotational splittings is highly non-trivial, an automated method for doing so has already been carried out by \cite{2015A&A...584A..50M} and subsequently put into practice \cite[e.g.][]{2016A&A...588A..87V}. Therefore, the presence of an extra, magnetic splitting of $1 \, \mu$Hz should be distinguishable with relative ease from rotational splittings. The results of our calculations for dipole and quadrupole modes are shown in tables \ref{tabledipole} and \ref{tablequadropole}, respectively, for the three chosen frequencies in each model. The notation $C_{\ell,m'} - C_{\ell,m}$ denotes the difference $\nu_{\ell, m'} - \nu_{\ell, m}$ and $B_\text{core,i}$, with $i = 1,2$, denotes the field value in the core that would make this difference equal to 1 $\mu$Hz. In what concerns the field configuration given by \ref{field2}, we chose the constant $c$ as the location of the edge of the radiative core, which occurs approximately at $0.1R$ for the 1.3\(M_\odot\) and 1.6\(M_\odot\) models and at $0.13R$ for the 2.0\(M_\odot\) model. As can be seen, for the dipole modes with frequencies closest to the estimated $\nu_\text{max}$, we find that magnetic fields with strengths $B \simeq 10^5$ G are necessary to produce a splitting of $1 \, \mu$Hz. This order of magnitude is the general trend for the other modes as well. Modes with increasing $|n|$ require lower field strengths to become split, and, as such, they are better suited to infer about the strength of the magnetic field in the core. This does not come as a surprise, as g-dominated mixed modes with lower frequencies have more of their inertia in the core and are more sensitive to the physical conditions of that region.

\begin{table}
\centering
        \begin{tabular}{c c c c c c}
            \toprule
             &  &  & \multicolumn{2}{c}{$C_{1,|1|} - C_{1,0}$} \\ \cline{4-5}
             & $|n|$ & $\nu \, (\mu$Hz) & $B_\text{core,1}$ (kG) & $B_\text{core,2}$ (kG) \\ \hline
            \multirow{3}{*}{$1.3$\(M_\odot\)} 
                                 & 100 & 118.41 & 27.1 & 27.2 \\
                                 & 89 & 130.16 & 31.2 & 31.4  \\ 
                                 & 80 & 141.40 & 35.4 & 35.6 \\ 
                                 \hline
            \multirow{3}{*}{$1.6$\(M_\odot\)} 
                                 & 108 & 113.66 & 25.3 & 25.4  \\ 
                                 & 98 & 123.00 & 28.1 & 28.3  \\ 
                                 & 88 & 133.89 & 34.7 & 34.8 \\ 
                                 \hline
            \multirow{3}{*}{$2.0$\(M_\odot\)} 
                                 & 90 & 84.77 & 52.0 & 52.2  \\ 
                                 & 80 & 93.07 & 59.8 & 60.0  \\ 
                                 & 70 & 102.71 & 70.1 & 70.3 \\
                                 \hline
        \end{tabular}
    \caption{The same as in table \ref{tabledipole} but for a frequency splitting of 11 nHz.}
    \label{tabledipoleminimum}
\end{table}

In the absence of observable magnetic splittings in the oscillation spectra of red giants, we can establish constraints to the core magnetic field by performing the same calculations for a frequency splitting below the frequency resolution, which is approximately $11.5$ nHz \citep{2012A&A...548A..10M}. With this in mind, we estimated the necessary field strength to produce a splitting of $11$ nHz in the frequencies of the same oscillation modes. The results for dipole and quadrupole modes are shown in tables \ref{tabledipoleminimum} and \ref{tablequadrupoleminimum}, respectively, and we find that the field has to be limited to a strength of the order of $10^4$ G. This value thus constitutes an upper limit to the core magnetic field in red giants in which no frequency splittings of magnetic origin are observed. 

\begin{table*}
\centering
        \begin{tabular}{c c c c c c c c c c c}
            \toprule
             &  &  & \multicolumn{2}{c}{$C_{2,0} - C_{2,|1|}$} & & \multicolumn{2}{c}{$C_{2,0} - C_{2,|2|}$} & & \multicolumn{2}{c}{$C_{2,|1|} - C_{2,|2|}$} \\ \cline{4-5} \cline{7-8} \cline{10-11} 
             & $|n|$ & $\nu \, (\mu$Hz) & $B_\text{core,1}$ (kG) &  $B_\text{core,2}$ (kG) & & $B_\text{core,1}$ (kG) &  $B_\text{core,2}$ (kG) & & $B_\text{core,1}$ (kG) &  $B_\text{core,2}$ (kG) \\ \hline 
            \multirow{3}{*}{$1.3$\(M_\odot\)} 
& 180 & 118.57 & 32.0 & 32.2 & & 16.0 & 16.1 & & 18.5 & 18.6 \\
& 162 & 130.12 & 37.0 & 37.3 & & 18.5 & 18.6 & & 21.3 & 21.5 \\
& 147 & 141.47 & 41.9 & 42.2 & & 21.0 & 21.1 & & 24.2 & 24.4 \\ \hline
            \multirow{3}{*}{$1.6$\(M_\odot\)} 
& 196 & 112.69 & 28.8 & 28.9  & & 14.4 & 14.5 & & 16.6 & 16.7 \\  
& 177 & 123.39 & 32.8 & 32.9 & & 16.4 & 16.5 & & 18.9 & 19.0 \\
& 162 & 133.24 & 36.9 & 37.0 & & 18.4 & 18.5 & & 21.3 & 21.4 \\
\hline
            \multirow{3}{*}{$2.0$\(M_\odot\)} 
& 163 & 84.75 & 61.2 & 61.4 & & 30.6 & 30.7 & & 35.3 & 35.4 \\ 
& 147 & 92.72 & 70.7 & 70.9 & & 35.3 & 35.5 & & 40.8 & 40.9 \\ 
& 132 & 101.65 & 80.3 & 80.6 & & 40.2 & 40.3 & & 46.4 & 46.5 \\ \hline
        \end{tabular}
    \caption{The same as in table \ref{tablequadropole} but for a frequency splitting of 11 nHz.}
    \label{tablequadrupoleminimum}
\end{table*}

\section{Discussion}
\label{sec:discussion}

The main goal of this work was to study magnetic fields in the radiative cores of evolved red giant stars using asteroseismology. Mixed oscillation modes in red giants are g-dominated, which means they can be used to probe the stellar core and obtain valuable information about the physical conditions of that region. Since magnetic fields are known to produce a splitting of the frequencies of the oscillation modes, magnetism deep in the interior of red giants could be inferred via this imprint.

We considered two distinct poloidal and axisymmetric field configurations in our analysis, which, along with the well-known perturbative approach, we used to derive an analytical expression for the magnetic frequency splitting of low angular degree and high absolute radial order g-dominated mixed modes. This expression shows effectively that the effect of the field is to lift the degeneracy imposed by spherical symmetry and split the frequencies of modes of angular degree  $\ell$ into $\ell + 1$ eigenmodes, as also found in the work of \cite{2005A&A...444L..29H} and \cite{2007MNRAS.377..453R} for g-modes in slowly pulsating B-stars and the Sun, respectively. As explained in the Subsection \ref{subsec:limit}, his analysis has some limitations, such as the disregard of the coupling of adjacent eigenmodes and the effect of the magnetic field on the propagation of gravity waves. Nevertheless, we believe that these limitations do not compromise our results.

Using the aforementioned expression, we proceeded to infer magnetic field strengths in the cores of red giant stars. We considered three red giant models with masses of 1.3\(M_\odot\), 1.6\(M_\odot\) and 2.0\(M_\odot\), which we believe are representative of the ensemble of evolved low-mass red giants observed by the Kepler space mission. Our calculations suggest that a central field strength of $10^5$ G should produce a splitting of the order of a $\mu$Hz in the frequencies of dipole and quadrupole oscillation modes. This splitting is in the observable range and is sufficiently distinguishable from rotational splittings. Moreover, we find that, in the absence of observable magnetic splittings, the magnetic field in the core must be limited to strengths of the order of $10^4$ G. This result is a constraint to the core magnetic field in red giants that do not exhibit frequency splittings of magnetic origin in their oscillation spectra. Although the considered configurations are purely poloidal and therefore unstable, and the ideal arrangement would be a field with both poloidal and toroidal components, we do not expect our conclusions to be greatly altered by the absence of the toroidal counterpart, as the magnetic frequency splitting is much more sensitive to the poloidal component.

In what concerns other studies of this nature, \cite{2015Sci...350..423F} and \cite{2016ApJ...824...14C} proposed the existence of strong fields $(B \gtrsim 10^5 G)$ in the cores of red giants showing depressed dipole modes \citep{2012A&A...537A..30M,2016Natur.529..364S} in order to explain this phenomenon. Although this is a very interesting and appealing scenario, according to our results a field strength of the order of $10^5$G should produce an observable frequency splitting, which was not observed experimentally in the analysis conducted by \cite{2017A&A...598A..62M}. In this work, the authors investigated a specific group of red giants with depressed but still measurable dipole modes, and found no evidence of magnetic splittings; in comparison with the mixed-mode pattern of a set of reference stars, which do not exhibit this feature, no significant differences apart from the mode amplitudes were observed. Therefore, under the assumption that the mechanism responsible for causing the mode suppression affects all azimuthal orders $m$ equally, our results combined with these observations suggest that fields of the order of $10^4 - 10^5$ G are not common throughout the cores of red giant stars.  Effectively, if this mechanism were to selectively remove some eigenmodes of a multiplet while keeping others, the failure to observe either could be interpreted as a total absence of splitting; thus, there is the need to assume that the mechanism is not biased towards some values of  $m$. This is a very possible scenario, given that the anisotropy of magnetic fields can perfectly lead to a suppression mechanism that depends on $m$, assuming, of course, that this mechanism is caused by magnetism and not by some other phenomenon. We hope that in the future, with the analysis of the data collected by the already on-going \textit{TESS} space mission \cite[e.g.][]{2015JATIS...1a4003R} and the forthcoming PLATO mission \cite[e.g.][]{2014ExA....38..249R}, experimental observations may shed some light on the subject of magnetism in the interior of stars, and allow this phenomenon to be better understood.

\section*{Acknowledgements}

We are grateful to the authors of MESA and GYRE for having made their codes publicly available, and to the authors \cite{2015Sci...350..423F} for making the MESA inlist used in their work available. We thank the Fun\-da\c c\~ao para a Ci\^encia e Tecnologia (FCT), Portugal, for the financial support to the Center for Astrophysics and Gravitation-CENTRA, Instituto Superior T\'ecnico, Universidade de Lisboa, through the Grant Project~No.~UIDB/00099/2020. We are also grateful to the anonymous referee for the helpful commentaries and suggestions.






\bibliographystyle{mnras}
\bibliography{mnras_astro-ph} 

\begin{thebibliography}{}
\makeatletter
\relax
\def\mn@urlcharsother{\let\do\@makeother \do\$\do\&\do\#\do\^\do\_\do\%\do\~}
\def\mn@doi{\begingroup\mn@urlcharsother \@ifnextchar [ {\mn@doi@}
  {\mn@doi@[]}}
\def\mn@doi@[#1]#2{\def\@tempa{#1}\ifx\@tempa\@empty \href
  {http://dx.doi.org/#2} {doi:#2}\else \href {http://dx.doi.org/#2} {#1}\fi
  \endgroup}
\def\mn@eprint#1#2{\mn@eprint@#1:#2::\@nil}
\def\mn@eprint@arXiv#1{\href {http://arxiv.org/abs/#1} {{\tt arXiv:#1}}}
\def\mn@eprint@dblp#1{\href {http://dblp.uni-trier.de/rec/bibtex/#1.xml}
  {dblp:#1}}
\def\mn@eprint@#1:#2:#3:#4\@nil{\def\@tempa {#1}\def\@tempb {#2}\def\@tempc
  {#3}\ifx \@tempc \@empty \let \@tempc \@tempb \let \@tempb \@tempa \fi \ifx
  \@tempb \@empty \def\@tempb {arXiv}\fi \@ifundefined
  {mn@eprint@\@tempb}{\@tempb:\@tempc}{\expandafter \expandafter \csname
  mn@eprint@\@tempb\endcsname \expandafter{\@tempc}}}

\bibitem[\protect\citeauthoryear{{Aerts}, {Christensen-Dalsgaard}  \&
  {Kurtz}}{{Aerts} et~al.}{2010}]{2010aste.book.....A}
{Aerts} C.,  {Christensen-Dalsgaard} J.,   {Kurtz} D.~W.,  2010,
  {Asteroseismology}.
{Springer Netherlands}

\bibitem[\protect\citeauthoryear{{Auri{\`e}re} et~al.,}{{Auri{\`e}re}
  et~al.}{2007}]{2007A&A...475.1053A}
{Auri{\`e}re} M.,  et~al., 2007, \aap, 475

\bibitem[\protect\citeauthoryear{{Auri{\`e}re} et~al.,}{{Auri{\`e}re}
  et~al.}{2015}]{2015A&A...574A..90A}
{Auri{\`e}re} M.,  et~al., 2015, \aap, 574, A90

\bibitem[\protect\citeauthoryear{{Babcock}}{{Babcock}}{1947}]{1947ApJ...105..105B}
{Babcock} H.~W.,  1947, \apj, 105, 105

\bibitem[\protect\citeauthoryear{{Baglin}, {Auvergne}, {Barge}, {Michel},
  {Catala}, {Deleuil}  \& {Weiss}}{{Baglin} et~al.}{2007}]{2007AIPC..895..201B}
{Baglin} A.,  {Auvergne} M.,  {Barge} P.,  {Michel} E.,  {Catala} C.,
  {Deleuil} M.,   {Weiss} W.,  2007, in {Dumitrache} C.,  {Popescu} N.~A.,
  {Suran} M.~D.,   {Mioc} V.,  eds,  American Institute of Physics Conference
  Series Vol. 895, Fifty Years of Romanian Astrophysics. pp 201--209

\bibitem[\protect\citeauthoryear{{Beck} et~al.,}{{Beck}
  et~al.}{2012}]{2012Natur.481...55B}
{Beck} P.~G.,  et~al., 2012, \nat, 481, 55

\bibitem[\protect\citeauthoryear{{Bedding} et~al.,}{{Bedding}
  et~al.}{2011}]{2011Natur.471..608B}
{Bedding} T.~R.,  et~al., 2011, \nat, 471, 608

\bibitem[\protect\citeauthoryear{{Belkacem}, {Goupil}, {Dupret}, {Samadi},
  {Baudin}, {Noels}  \& {Mosser}}{{Belkacem}
  et~al.}{2011}]{2011A&A...530A.142B}
{Belkacem} K.,  {Goupil} M.~J.,  {Dupret} M.~A.,  {Samadi} R.,  {Baudin} F.,
  {Noels} A.,   {Mosser} B.,  2011, \aap, 530, A142

\bibitem[\protect\citeauthoryear{{Biront}, {Goossens}, {Cousens}  \&
  {Mestel}}{{Biront} et~al.}{1982}]{1982MNRAS.201..619B}
{Biront} D.,  {Goossens} M.,  {Cousens} A.,   {Mestel} L.,  1982, \mnras, 201,
  619

\bibitem[\protect\citeauthoryear{{Braithwaite}}{{Braithwaite}}{2007}]{2007A&A...469..275B}
{Braithwaite} J.,  2007, \aap, 469, 275

\bibitem[\protect\citeauthoryear{{Braithwaite} \& {Nordlund}}{{Braithwaite} \&
  {Nordlund}}{2006}]{2006A&A...450.1077B}
{Braithwaite} J.,  {Nordlund} {\r{A}}.,  2006, \aap, 450, 1077

\bibitem[\protect\citeauthoryear{{Braithwaite} \& {Spruit}}{{Braithwaite} \&
  {Spruit}}{2004}]{2004Natur.431..819B}
{Braithwaite} J.,  {Spruit} H.~C.,  2004, \nat, 431, 819

\bibitem[\protect\citeauthoryear{{Braithwaite} \& {Spruit}}{{Braithwaite} \&
  {Spruit}}{2017}]{2017RSOS....460271B}
{Braithwaite} J.,  {Spruit} H.~C.,  2017, Royal Society Open Science, 4, 160271

\bibitem[\protect\citeauthoryear{{Busso}, {Wasserburg}, {Nollett}  \&
  {Calandra}}{{Busso} et~al.}{2007}]{2007ApJ...671..802B}
{Busso} M.,  {Wasserburg} G.~J.,  {Nollett} K.~M.,   {Calandra} A.,  2007,
  \apj, 671, 802

\bibitem[\protect\citeauthoryear{{Cantiello} \& {Braithwaite}}{{Cantiello} \&
  {Braithwaite}}{2011}]{2011A&A...534A.140C}
{Cantiello} M.,  {Braithwaite} J.,  2011, \aap, 534, A140

\bibitem[\protect\citeauthoryear{{Cantiello}, {Fuller}  \&
  {Bildsten}}{{Cantiello} et~al.}{2016}]{2016ApJ...824...14C}
{Cantiello} M.,  {Fuller} J.,   {Bildsten} L.,  2016, \apj, 824, 14

\bibitem[\protect\citeauthoryear{{Chaplin} \& {Miglio}}{{Chaplin} \&
  {Miglio}}{2013}]{2013ARA&A..51..353C}
{Chaplin} W.~J.,  {Miglio} A.,  2013, \araa, 51, 353

\bibitem[\protect\citeauthoryear{{Chaplin} et~al.,}{{Chaplin}
  et~al.}{2011}]{2011Sci...332..213C}
{Chaplin} W.~J.,  et~al., 2011, Science, 332, 213

\bibitem[\protect\citeauthoryear{{Donati} \& {Landstreet}}{{Donati} \&
  {Landstreet}}{2009}]{2009ARA&A..47..333D}
{Donati} J.~F.,  {Landstreet} J.~D.,  2009, \araa, 47, 333

\bibitem[\protect\citeauthoryear{{Donati}, {Semel}, {Carter}, {Rees}  \&
  {Collier Cameron}}{{Donati} et~al.}{1997}]{1997MNRAS.291..658D}
{Donati} J.~F.,  {Semel} M.,  {Carter} B.~D.,  {Rees} D.~E.,   {Collier
  Cameron} A.,  1997, \mnras, 291, 658

\bibitem[\protect\citeauthoryear{{Duez} \& {Mathis}}{{Duez} \&
  {Mathis}}{2010}]{2010A&A...517A..58D}
{Duez} V.,  {Mathis} S.,  2010, \aap, 517, A58

\bibitem[\protect\citeauthoryear{{Duez}, {Braithwaite}  \& {Mathis}}{{Duez}
  et~al.}{2010}]{2010ApJ...724L..34D}
{Duez} V.,  {Braithwaite} J.,   {Mathis} S.,  2010, \apjl, 724, L34

\bibitem[\protect\citeauthoryear{{Dziembowski} \& {Goode}}{{Dziembowski} \&
  {Goode}}{1985}]{1985ApJ...296L..27D}
{Dziembowski} W.,  {Goode} P.~R.,  1985, \apjl, 296, L27

\bibitem[\protect\citeauthoryear{{Eckart}}{{Eckart}}{1960}]{1960hydro.book.....F}
{Eckart} C.,  1960, {Hydrodynamics of Oceans and Atmospheres}.
{ Pergammon Press, Oxford}

\bibitem[\protect\citeauthoryear{{Fuller}, {Cantiello}, {Stello}, {Garcia}  \&
  {Bildsten}}{{Fuller} et~al.}{2015}]{2015Sci...350..423F}
{Fuller} J.,  {Cantiello} M.,  {Stello} D.,  {Garcia} R.~A.,   {Bildsten} L.,
  2015, Science, 350, 423

\bibitem[\protect\citeauthoryear{{Gilliland} et~al.,}{{Gilliland}
  et~al.}{2010}]{2010PASP..122..131G}
{Gilliland} R.~L.,  et~al., 2010, \pasp, 122, 131

\bibitem[\protect\citeauthoryear{{Gough} \& {Taylor}}{{Gough} \&
  {Taylor}}{1984}]{1984MmSAI..55..215G}
{Gough} D.~O.,  {Taylor} P.~P.,  1984, \memsai, 55, 215

\bibitem[\protect\citeauthoryear{{Gough} \& {Thompson}}{{Gough} \&
  {Thompson}}{1990}]{1990MNRAS.242...25G}
{Gough} D.~O.,  {Thompson} M.~J.,  1990, \mnras, 242, 25

\bibitem[\protect\citeauthoryear{{Goupil}, {Mosser}, {Marques}, {Ouazzani},
  {Belkacem}, {Lebreton}  \& {Samadi}}{{Goupil}
  et~al.}{2013}]{2013A&A...549A..75G}
{Goupil} M.~J.,  {Mosser} B.,  {Marques} J.~P.,  {Ouazzani} R.~M.,  {Belkacem}
  K.,  {Lebreton} Y.,   {Samadi} R.,  2013, \aap, 549, A75

\bibitem[\protect\citeauthoryear{{Hasan}, {Zahn}  \&
  {Christensen-Dalsgaard}}{{Hasan} et~al.}{2005}]{2005A&A...444L..29H}
{Hasan} S.~S.,  {Zahn} J.~P.,   {Christensen-Dalsgaard} J.,  2005, \aap, 444,
  L29

\bibitem[\protect\citeauthoryear{{Jones}, {Pesnell}, {Hansen}  \&
  {Kawaler}}{{Jones} et~al.}{1989}]{1989ApJ...336..403J}
{Jones} P.~W.,  {Pesnell} W.~D.,  {Hansen} C.~J.,   {Kawaler} S.~D.,  1989,
  \apj, 336, 403

\bibitem[\protect\citeauthoryear{{Kamchatnov}}{{Kamchatnov}}{1982}]{2004physics...9093K}
{Kamchatnov} A.~M.,  1982, Zh. Eksp. Teor. Fiz, p.~117

\bibitem[\protect\citeauthoryear{{Kiefer} \& {Roth}}{{Kiefer} \&
  {Roth}}{2018}]{2018ApJ...854...74K}
{Kiefer} R.,  {Roth} M.,  2018, \apj, 854, 74

\bibitem[\protect\citeauthoryear{{Kiefer}, {Schad}  \& {Roth}}{{Kiefer}
  et~al.}{2017}]{2017ApJ...846..162K}
{Kiefer} R.,  {Schad} A.,   {Roth} M.,  2017, \apj, 846, 162

\bibitem[\protect\citeauthoryear{{Konstantinova-Antova}
  et~al.,}{{Konstantinova-Antova} et~al.}{2014}]{2014IAUS..302..373K}
{Konstantinova-Antova} R.,  et~al., 2014, in {Petit} P.,  {Jardine} M.,
  {Spruit} H.~C.,  eds,  IAU Symposium Vol. 302, Magnetic Fields throughout
  Stellar Evolution. pp 373--376

\bibitem[\protect\citeauthoryear{{Lagarde}, {Bossini}, {Miglio}, {Vrard}  \&
  {Mosser}}{{Lagarde} et~al.}{2016}]{2016MNRAS.457L..59L}
{Lagarde} N.,  {Bossini} D.,  {Miglio} A.,  {Vrard} M.,   {Mosser} B.,  2016,
  \mnras, 457, L59

\bibitem[\protect\citeauthoryear{{Lavely} \& {Ritzwoller}}{{Lavely} \&
  {Ritzwoller}}{1992}]{1992RSPTA.339..431L}
{Lavely} E.~M.,  {Ritzwoller} M.~H.,  1992, Philosophical Transactions of the
  Royal Society of London Series A, 339, 431

\bibitem[\protect\citeauthoryear{{Lecoanet}, {Vasil}, {Fuller}, {Cantiello}  \&
  {Burns}}{{Lecoanet} et~al.}{2017}]{2017MNRAS.466.2181L}
{Lecoanet} D.,  {Vasil} G.~M.,  {Fuller} J.,  {Cantiello} M.,   {Burns} K.~J.,
  2017, \mnras, 466, 2181

\bibitem[\protect\citeauthoryear{{Ledoux} \& {Simon}}{{Ledoux} \&
  {Simon}}{1957}]{1957AnAp...20..185L}
{Ledoux} P.,  {Simon} R.,  1957, Annales d'Astrophysique, 20, 185

\bibitem[\protect\citeauthoryear{{Ligni{\`e}res}, {Petit}, {B{\"o}hm}  \&
  {Auri{\`e}re}}{{Ligni{\`e}res} et~al.}{2009}]{2009A&A...500L..41L}
{Ligni{\`e}res} F.,  {Petit} P.,  {B{\"o}hm} T.,   {Auri{\`e}re} M.,  2009,
  \aap, 500, L41

\bibitem[\protect\citeauthoryear{{Loi}}{{Loi}}{2020}]{2020MNRAS.493.5726L}
{Loi} S.~T.,  2020, \mnras, 493, 5726

\bibitem[\protect\citeauthoryear{{Loi} \& {Papaloizou}}{{Loi} \&
  {Papaloizou}}{2017}]{2017MNRAS.467.3212L}
{Loi} S.~T.,  {Papaloizou} J. C.~B.,  2017, \mnras, 467, 3212

\bibitem[\protect\citeauthoryear{{Loi} \& {Papaloizou}}{{Loi} \&
  {Papaloizou}}{2018}]{2018MNRAS.477.5338L}
{Loi} S.~T.,  {Papaloizou} J. C.~B.,  2018, \mnras, 477, 5338

\bibitem[\protect\citeauthoryear{{Markey} \& {Tayler}}{{Markey} \&
  {Tayler}}{1973}]{1973MNRAS.163...77M}
{Markey} P.,  {Tayler} R.~J.,  1973, \mnras, 163, 77

\bibitem[\protect\citeauthoryear{{Mosser} et~al.,}{{Mosser}
  et~al.}{2012a}]{2012A&A...537A..30M}
{Mosser} B.,  et~al., 2012a, \aap, 537, A30

\bibitem[\protect\citeauthoryear{{Mosser} et~al.,}{{Mosser}
  et~al.}{2012b}]{2012A&A...548A..10M}
{Mosser} B.,  et~al., 2012b, \aap, 548, A10

\bibitem[\protect\citeauthoryear{{Mosser}, {Vrard}, {Belkacem}, {Deheuvels}  \&
  {Goupil}}{{Mosser} et~al.}{2015}]{2015A&A...584A..50M}
{Mosser} B.,  {Vrard} M.,  {Belkacem} K.,  {Deheuvels} S.,   {Goupil} M.~J.,
  2015, \aap, 584, A50

\bibitem[\protect\citeauthoryear{{Mosser} et~al.,}{{Mosser}
  et~al.}{2017}]{2017A&A...598A..62M}
{Mosser} B.,  et~al., 2017, \aap, 598, A62

\bibitem[\protect\citeauthoryear{{Nordhaus}, {Busso}, {Wasserburg}, {Blackman}
  \& {Palmerini}}{{Nordhaus} et~al.}{2008}]{2008ApJ...684L..29N}
{Nordhaus} J.,  {Busso} M.,  {Wasserburg} G.~J.,  {Blackman} E.~G.,
  {Palmerini} S.,  2008, \apjl, 684, L29

\bibitem[\protect\citeauthoryear{{Osaki}}{{Osaki}}{1975}]{1975PASJ...27..237O}
{Osaki} Y.,  1975, \pasj, 27, 237

\bibitem[\protect\citeauthoryear{{Paxton}, {Bildsten}, {Dotter}, {Herwig},
  {Lesaffre}  \& {Timmes}}{{Paxton} et~al.}{2011}]{2011ApJS..192....3P}
{Paxton} B.,  {Bildsten} L.,  {Dotter} A.,  {Herwig} F.,  {Lesaffre} P.,
  {Timmes} F.,  2011, \apjs, 192, 3

\bibitem[\protect\citeauthoryear{{Paxton} et~al.,}{{Paxton}
  et~al.}{2013}]{2013ApJS..208....4P}
{Paxton} B.,  et~al., 2013, \apjs, 208, 4

\bibitem[\protect\citeauthoryear{{Paxton} et~al.,}{{Paxton}
  et~al.}{2015}]{2015ApJS..220...15P}
{Paxton} B.,  et~al., 2015, \apjs, 220, 15

\bibitem[\protect\citeauthoryear{{Paxton} et~al.,}{{Paxton}
  et~al.}{2018}]{2018ApJS..234...34P}
{Paxton} B.,  et~al., 2018, \apjs, 234, 34

\bibitem[\protect\citeauthoryear{{Paxton} et~al.,}{{Paxton}
  et~al.}{2019}]{2019ApJS..243...10P}
{Paxton} B.,  et~al., 2019, \apjs, 243, 10

\bibitem[\protect\citeauthoryear{{Prat}, {Mathis}, {Buysschaert}, {Van Beeck},
  {Bowman}, {Aerts}  \& {Neiner}}{{Prat} et~al.}{2019}]{2019A&A...627A..64P}
{Prat} V.,  {Mathis} S.,  {Buysschaert} B.,  {Van Beeck} J.,  {Bowman} D.~M.,
  {Aerts} C.,   {Neiner} C.,  2019, \aap, 627, A64

\bibitem[\protect\citeauthoryear{{Prat}, {Mathis}, {Neiner}, {Van Beeck},
  {Bowman}  \& {Aerts}}{{Prat} et~al.}{2020}]{2020A&A...636A.100P}
{Prat} V.,  {Mathis} S.,  {Neiner} C.,  {Van Beeck} J.,  {Bowman} D.~M.,
  {Aerts} C.,  2020, \aap, 636, A100

\bibitem[\protect\citeauthoryear{{Rashba}, {Semikoz}, {Turck-Chi{\`e}ze}  \&
  {Valle}}{{Rashba} et~al.}{2007}]{2007MNRAS.377..453R}
{Rashba} T.~I.,  {Semikoz} V.~B.,  {Turck-Chi{\`e}ze} S.,   {Valle} J.~W.~F.,
  2007, \mnras, 377, 453

\bibitem[\protect\citeauthoryear{{Rauer} et~al.,}{{Rauer}
  et~al.}{2014}]{2014ExA....38..249R}
{Rauer} H.,  et~al., 2014, Experimental Astronomy, 38, 249

\bibitem[\protect\citeauthoryear{{Ricker} et~al.,}{{Ricker}
  et~al.}{2015}]{2015JATIS...1a4003R}
{Ricker} G.~R.,  et~al., 2015, Journal of Astronomical Telescopes, Instruments,
  and Systems, 1, 014003

\bibitem[\protect\citeauthoryear{{Roberts}}{{Roberts}}{1981}]{1981AN....302...65R}
{Roberts} P.~H.,  1981, Astronomische Nachrichten, 302, 65

\bibitem[\protect\citeauthoryear{{Roberts} \& {Soward}}{{Roberts} \&
  {Soward}}{1983}]{1983MNRAS.205.1171R}
{Roberts} P.~H.,  {Soward} A.~M.,  1983, \mnras, 205, 1171

\bibitem[\protect\citeauthoryear{{Scuflaire}}{{Scuflaire}}{1974}]{1974A&A....36..107S}
{Scuflaire} R.,  1974, \aap, 36, 107

\bibitem[\protect\citeauthoryear{{Serenelli} et~al.,}{{Serenelli}
  et~al.}{2017}]{2017ApJS..233...23S}
{Serenelli} A.,  et~al., 2017, \apjs, 233, 23

\bibitem[\protect\citeauthoryear{{Shibahashi} \& {Aerts}}{{Shibahashi} \&
  {Aerts}}{2000}]{2000ApJ...531L.143S}
{Shibahashi} H.,  {Aerts} C.,  2000, \apjl, 531, L143

\bibitem[\protect\citeauthoryear{{Stello}, {Chaplin}, {Basu}, {Elsworth}  \&
  {Bedding}}{{Stello} et~al.}{2009}]{2009MNRAS.400L..80S}
{Stello} D.,  {Chaplin} W.~J.,  {Basu} S.,  {Elsworth} Y.,   {Bedding} T.~R.,
  2009, \mnras, 400, L80

\bibitem[\protect\citeauthoryear{{Stello}, {Cantiello}, {Fuller}, {Huber},
  {Garc{\'\i}a}, {Bedding}, {Bildsten}  \& {Silva Aguirre}}{{Stello}
  et~al.}{2016}]{2016Natur.529..364S}
{Stello} D.,  {Cantiello} M.,  {Fuller} J.,  {Huber} D.,  {Garc{\'\i}a} R.~A.,
  {Bedding} T.~R.,  {Bildsten} L.,   {Silva Aguirre} V.,  2016, \nat, 529, 364

\bibitem[\protect\citeauthoryear{{Takata}}{{Takata}}{2006}]{2006PASJ...58..893T}
{Takata} M.,  2006, \pasj, 58, 893

\bibitem[\protect\citeauthoryear{{Tayler}}{{Tayler}}{1973}]{1973MNRAS.161..365T}
{Tayler} R.~J.,  1973, \mnras, 161, 365

\bibitem[\protect\citeauthoryear{{Townsend} \& {Teitler}}{{Townsend} \&
  {Teitler}}{2013}]{2013MNRAS.435.3406T}
{Townsend} R.~H.~D.,  {Teitler} S.~A.,  2013, \mnras, 435, 3406

\bibitem[\protect\citeauthoryear{{Unno}, {Osaki}, {Ando}, {Saio}  \&
  {Shibahashi}}{{Unno} et~al.}{1989}]{1989nos..book.....U}
{Unno} W.,  {Osaki} Y.,  {Ando} H.,  {Saio} H.,   {Shibahashi} H.,  1989,
  {Nonradial oscillations of stars}.
{University of Tokyo Press}

\bibitem[\protect\citeauthoryear{{Vrard}, {Mosser}  \& {Samadi}}{{Vrard}
  et~al.}{2016}]{2016A&A...588A..87V}
{Vrard} M.,  {Mosser} B.,   {Samadi} R.,  2016, \aap, 588, A87

\bibitem[\protect\citeauthoryear{{Wright}}{{Wright}}{1973}]{1973MNRAS.162..339W}
{Wright} G.~A.~E.,  1973, \mnras, 162, 339

\makeatother
\end{thebibliography}




\appendix

\section{The perturbed magnetic field}

In this appendix, we discuss in more detail the components of the perturbed magnetic field, expression \ref{perturbed}, and explain our reasoning behind the derivation of expression \ref{splittingsimple}.

In spherical coordinates $(r, \theta, \phi)$, the perturbed magnetic field can be written \citep[][]{2007MNRAS.377..453R}

\begin{equation}
\begin{split}
  \boldsymbol{B}' & (r, \theta, \phi, t) = \left[ (\boldsymbol{B} \cdot \nabla) \bar\xi_r - (\bm{\xi} \cdot \nabla)B_r - B_r \, (\nabla \cdot \bm{\xi}) \right] \mathbf{e}_r \\ & + \left[ (\boldsymbol{B} \cdot \nabla) \bar\xi_\theta - (\bm{\xi} \cdot \nabla)B_\theta + \frac{B_\theta \bar\xi_r - \bar\xi_\theta B_r}{r} - B_\theta \, (\nabla \cdot \bm{\xi}) \right] \mathbf{e}_\theta \\
  & + \bigg[ (\boldsymbol{B} \cdot \nabla) \bar\xi_\phi - (\bm{\xi} \cdot \nabla)B_\phi + \frac{B_\phi \bar\xi_r - \bar\xi_\phi B_r}{r} \\ & + \cot \theta \frac{(B_\phi \bar\xi_\theta - \bar\xi_\phi B_\theta)}{r} - B_\phi \, (\nabla \cdot \bm{\xi}) \bigg] \mathbf{e}_\phi 
\end{split}
\label{pertfield}
\end{equation}

where, as mentioned in the main text, $\boldsymbol{B}$ is the magnetic field vector, $\bm{\xi}$ is the displacement vector, expression \ref{displacement}, and

\begin{equation}
    \bar\xi_r (r, \theta, \phi, t) = \xi_r(r) \, Y^\ell_m \, \text{e}^{i \omega t} \, ,
\end{equation}
\begin{equation}
    \bar\xi_\theta (r, \theta, \phi, t) = \xi_h(r) \, \frac{\partial Y^\ell_m}{\partial \theta} \, \text{e}^{i \omega t} \, ,
\end{equation}
\begin{equation}
    \bar\xi_\phi (r, \theta, \phi, t) = \xi_h(r) \, \frac{im}{\sin \theta} \, Y^\ell_m \, \text{e}^{i \omega t} \, ,
\end{equation}

where $\xi_r$ and $\xi_h$ are, respectively, the radial and horizontal displacement functions, the functions $Y^\ell_m = Y^\ell_m(\theta, \phi)$ are the spherical harmonic functions of angular degree $\ell$ and azimuthal order $m$, $\omega$ is the angular frequency, and $t$ is the time coordinate.

In this work, we considered two distinct poloidal field configurations for the magnetic field in the stellar core, expressions \ref{field1} and \ref{field2}, illustrated, respectively, on the left and right of Fig. \ref{magnetic1}. These fields are of the form
\begin{equation}
    \boldsymbol{B} = B_0 \left( f(x) \, \cos \theta, g(x) \sin \theta, 0
    \right) \, ,
    \label{gen_field} 
\end{equation}

where $B_0$ is a constant, $x = r/R$ is the normalised radial coordinate, $R$ is the stellar radius, and $g(x) = (-1/2x) \partial (x^2 \, f(x))/ \partial x$. For the first field, $f(x) = (5 - 3x^2)/2$, whereas for the second $f(x) = 1/(1+(x/c)^2)^2$, where $c$ is a constant. Plugging the previous expression in \ref{pertfield} yields the following components,

\begin{equation}
    \begin{split}
    B_r' = \frac{B_0}{R} \bigg[ & \frac{\xi_h(x)}{x} f(x) \left( \ell(\ell+1) \, \cos \theta \, Y^\ell_m + \sin \theta \, \frac{\partial Y^\ell_m}{\partial \theta} \right) \\ & - \frac{\xi_r(x)}{2x^2 \sin \theta} \frac{\partial (\sin^2 \theta Y^\ell_m)}{\partial \theta} \frac{\partial (x^2 f(x))}{\partial x} \bigg] \, \text{e}^{i \omega t}
    \end{split} \, ,
    \label{bpr} 
\end{equation}

\begin{equation}
    \begin{split}
    B_\theta' = \frac{B_0}{R} \bigg[ & \cos \theta \, \frac{\partial Y^\ell_m}{\partial \theta} \frac{1}{x} \frac{\partial }{\partial x} (x \, f(x) \, \xi_h(x) ) \\ & - \sin \theta \, Y^\ell_m \frac{1}{2x} \frac{\partial }{\partial x} \left( \xi_r(x) \frac{\partial}{\partial x} (x^2 \, f(x) ) \right) \\ & - \frac{m^2}{\sin \theta} Y^\ell_m \frac{\xi_h(x)}{2x^2} \frac{\partial (x^2 \, f(x))}{\partial x} \bigg] \, \text{e}^{i \omega t}
    \end{split} \, ,
    \label{bptheta}
\end{equation}

\begin{equation}
    \begin{split}
    B_\phi' = i m \frac{B_0}{R} \bigg[ & \cot \theta \, Y^\ell_m \frac{1}{x} \frac{\partial}{\partial x} (x \, f(x) \, \xi_h(x)) \\ & - \xi_h(x) \frac{\partial Y^\ell_m}{\partial \theta} \frac{1}{2x^2} \frac{\partial}{\partial x} (x^2 \, f(x))  \bigg] \, \text{e}^{i \omega t} \, .
    \end{split} 
    \label{bpphi}
\end{equation}

As was pointed out extensively throughout the main text, red giant stars possess gravity-dominated mixed modes, and therefore the displacement from the equilibrium configuration occurs mostly in the horizontal direction. Additionally, since these modes have very high absolute radial orders ($|n| \gg 0$), in the perturbed magnetic field the terms involving the derivatives of the horizontal displacement $\xi_h$ are very large. This is explained in Section \ref{sec:asymptotic_expansion}, using analytical expressions for $\xi_h$ and $\xi_r$, obtained via the JWKB approximation, and in Section \ref{sec:model}. Such terms are present only in the angular components of the perturbed field, $B_\theta'$ and $B_\phi'$, where they appear multiplied by $\cos \theta \, (\partial Y^\ell_m/\partial \theta)$ and $\cot \theta \, Y^\ell_m$, respectively. In these conditions, the first term in the integral of expression \ref{splitting} dominates over the other two, and it can be written

\begin{equation}
\begin{split}
    |\boldsymbol{B}'|^2 \simeq & \left( \frac{B_0}{R} \right )^2 \left| \cos \theta \, \frac{\partial Y^\ell_m}{\partial \theta} \frac{1}{x} \frac{\partial }{\partial x} (x \, f(x) \, \xi_h(x) ) \right|^2 \\ + & \left( m \frac{B_0}{R} \right )^2 \left| \cot \theta \, Y^\ell_m \frac{1}{x} \frac{\partial}{\partial x} (x \, f(x) \, \xi_h(x)) \right|^2 \\
    = & \left( \frac{B_0}{R} \right)^2 \left| \frac{1}{x} \frac{\partial }{\partial x} (x \, f(x) \, \xi_h(x)) \right|^2 \Bigg( \left| \cos \theta \frac{\partial Y^\ell_m}{\partial \theta} \right|^2 \\ & +  m^2 \left| \cot \theta \, Y^\ell_m \right|^2  \Bigg)
\end{split}
\end{equation}

as in expression \ref{eqn1}. The second and third terms in the integral of expression \ref{splitting} can be neglected, as they scale, respectively, with $\partial \xi_h/\partial x$ and $\xi_h$. In addition, since $|\xi_h| \gg |\xi_r|$, the main contribution to the mode inertia, expression \ref{mode_inertia}, is the term containing $\xi_h^2$, and the overall expression for the splitting \ref{splitting} can be written

\begin{equation}
    \frac{\delta \omega}{\omega} = \frac{1}{8 \pi \omega^2} \left( \frac{B_0}{R} \right)^2 \, \mathcal{I} \, C_{\ell,m}
\end{equation}

where

\begin{equation}
    \mathcal{I} = \frac{\int \left| \frac{1}{x} \frac{\partial }{\partial x} (x \, f(x) \, \xi_h) \right|^2 \, x^2 dx}{ \ell(\ell+1) \int \xi_h^2 \, \rho \, x^2 \, \text{d}x} 
\end{equation}

\begin{equation}
    C_{\ell,m} = \int \left( \left| \cos \theta \frac{\partial Y^\ell_m}{\partial \theta} \right|^2 + m^2 \left| \cot \theta \, Y^\ell_m \right|^2 \right) \, \sin \theta \, d\theta.
\end{equation}

These expressions are very similar to the ones found in the works of \cite{2005A&A...444L..29H} and \cite{2007MNRAS.377..453R}, for the case of gravity modes in slowly pulsating B-stars and in the Sun, respectively.


\bsp	
\label{lastpage}
\end{document}